# An ultrasonic transducer for vibration mode conversion of wedge-shaped structure of acoustic black hole*

Yi Wang, Cheng Chen, Shuyu Lin*

Shaanxi Key Laboratory of Ultrasonics, Institute of Applied Acoustics, Shaanxi Normal University,
Xi'an 710119, China

*sylin@snnu.edu.cn

**Abstract**

Acoustic black hole (ABH) structure has been extensively employed in applications such as vibration mitigation, noise reduction, and energy harvesting, owing to its unique sound wave trapping and energy concentration effects. Furthermore, ABH structure shows significant promise in improving the performance of ultrasonic device and constructing multifunctional acoustic field. Therefore, this paper proposes an ultrasonic mode-conversion transducer consisting of a Langevin transducer and an ABH wedge radiant plate to investigate the potential applications of ABH in ultrasonic levitation and multifunctional particle manipulation. The theoretical model of flexural vibration of the radiant plate was established by utilizing Timoshenko beam theory and transfer matrix method, and the calculated vibration frequencies demonstrated good agreement with those obtained from finite element simulations (FES). The electrical impedance frequency response characteristics, vibration modes and the near-field sound pressure distribution of the transducer in air were also simulated. The results revealed that the amplitude of the ABH wedge radiant plate increases stepwise, and the sound pressure exhibits a gradient distribution. A prototype of the transducer was fabricated and experimentally tested, confirming the accuracy of FES and the feasibility of the design approach. Finally, the ultrasonic levitation experiment demonstrated that the ABH design enables the formation of gradient distribution of sound pressure in the standing wave sound field, thereby facilitating precise particle sorting.

## 1. Introduction

In recent years, acoustic black hole structures have garnered considerable attention as a novel class of structures for vibration and noise reduction. An ABH structure is characterized by a special geometric design with variable thickness, which effectively decelerates flexural waves. As the thickness approaches zero, the flexural wave velocity also approaches zero, leading to an infinite accumulation of phase and amplitude. This phenomenon causes flexural waves to concentrate at the tip of the structure, thereby demonstrating sound wave trapping and energy concentration effects of the ABH structure. Even with a practical truncation thickness, ABH structures still exhibit considerable sound wave trapping and energy concentration capabilities[1-4]. Consequently, ABH structures have found widespread application across various fields. Beyond vibration and noise reduction[5-7], energy harvesting[8-12], significant advancements have been made in improving device performance. For instance, Remillieux et al.[13, 14] incorporated ABH structure into the design of stacked piezoelectric transducer, achieving more efficient air radiation. Liu et al.[15] embedded the ABH structure into the design of radial composite transducer, resulting in improved electromechanical conversion and acoustic radiation performance, alongside enhanced radiation directivity. Chen et al.[16] designed a novel ABH immersed sonoreactor, which notably expanded the ultrasonic cavitation region and intensified acoustic radiation, thus greatly improving the efficiency of sonochemical treatment. In the realm of multifunctional particle manipulation, ABH structure has also shown promise. Liu et al.[17] deposited a polystyrene-microparticle suspension into an ABH depression on an elliptical polymethyl-methacrylate plate, and harnessed the acoustic streaming generated by the two-dimensional ABH structure to achieve the enrichment and patterning of polystyrene particles. Yin et al.[18] achieved the capture, transfer, and patterning of acrylic particles by employing an ABH ultrasonic microprobe to stimulate acoustic streaming in liquid. Collectively, these studies underscore that ABH structures can significantly enhance device radiation capabilities, thereby holding substantial potential for multi-functional acoustic field construction and particle manipulation.

Based on the sound wave trapping and energy concentration effects of ABH structure, this paper introduces a novel vibration mode conversion ultrasonic transducer incorporating an ABH wedge structure. A theoretical analysis model for the flexural vibration of the radiant plate was established, and the calculated vibration frequencies were validated against those obtained from finite element simulation (FES). The electrical impedance frequency response characteristics, vibration modes, and near-field sound pressure distribution of the transducer in air were also simulated. A prototype of the transducer was fabricated, and experiments including impedance analysis, laser vibration measurement and sound field measurement in air were conducted. The experimental results were then compared with the FES results to evaluate the vibration and radiation performance of the transducer. Finally, an ultrasonic levitation experiment was performed to demonstrate the potential of the ABH transducer for constructing a multi-functional sound field. This study aims to investigate the potential applications of ABH structure in ultrasonic levitation technology, offering both theoretical methodologies and experimental reference for device design.

## 2. Theoretical modeling

The geometric structure of the ABH transducer is presented in Figs. 1 and 2. In Fig. 1, the gray part is made of steel, the silvery-white part is aluminum, and the yellow part is PZT-4 piezoelectric ceramic. Fig. 2(a) provides the geometric dimensions of the Langevin transducer. In Fig. 2(b), $h_1$ is the thickness of the uniform-thickness section of the ABH radiant plate, $h_0$ represent the truncation thickness, $l_{abh}$ donates the length of the ABH section, $w$ is the width of the radiant plate, $l$ is the total length of the radiant plate, and $l_b$ is the length of the uniform-thickness section. The exponent is denoted by $m$, and the ABH contour curve adheres to a power law described by the accompanying equation.

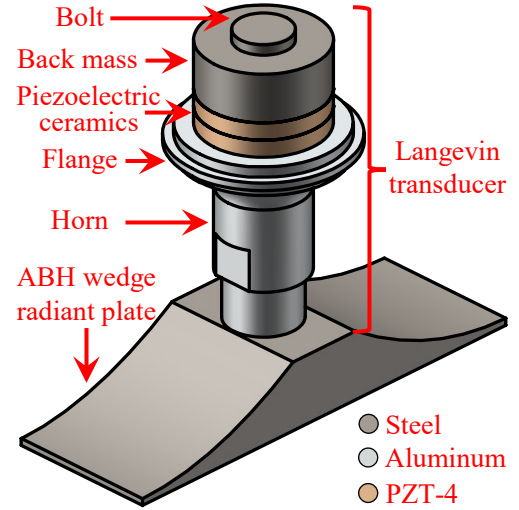

Fig. 1. Ultrasonic transducer with acoustic black hole wedge structure based on vibration mode-conversion.

According to the resonance system design principle, the first-order longitudinal vibration of the Langevin transducer was matched to the same frequency as the flexural vibration frequency of the radiant plate[19]. Based on the general design theory of Langevin transducer, the Langevin transducer depicted in Fig. 1 and Fig. 2(a) was designed, with its first-order longitudinal vibration frequency measured at 26055 Hz[20]. Given the weak coupling degree of the low-order one-dimensional flexural vibration modes of the radiant plate, the Timoshenko beam theory was employed to establish a transfer matrix theoretical model for the ABH wedge radiant plate. This approach enables rapid design of the radiant plate with a symmetrical flexural vibration frequency close to 26055 Hz.

### 2.1. Transfer matrix model based on Timoshenko beam theory

First, the variable-thickness section of the radiant plate needs to be discretized into several equally-spaced vibration elements. As illustrated in Fig. 3, both the left and right ABH sections are divided into $n$ vibration elements, with the uniform-thickness portion considered as a single vibration element[21]. The $x$-coordinate denotes the length direction, and the $y$-coordinate represents the thickness direction.

According to the Timoshenko beam theory, the flexural transverse displacement $Y_b$ and shear transverse displacement $Y_s$ of the radiant plate are defined as[22, 23]

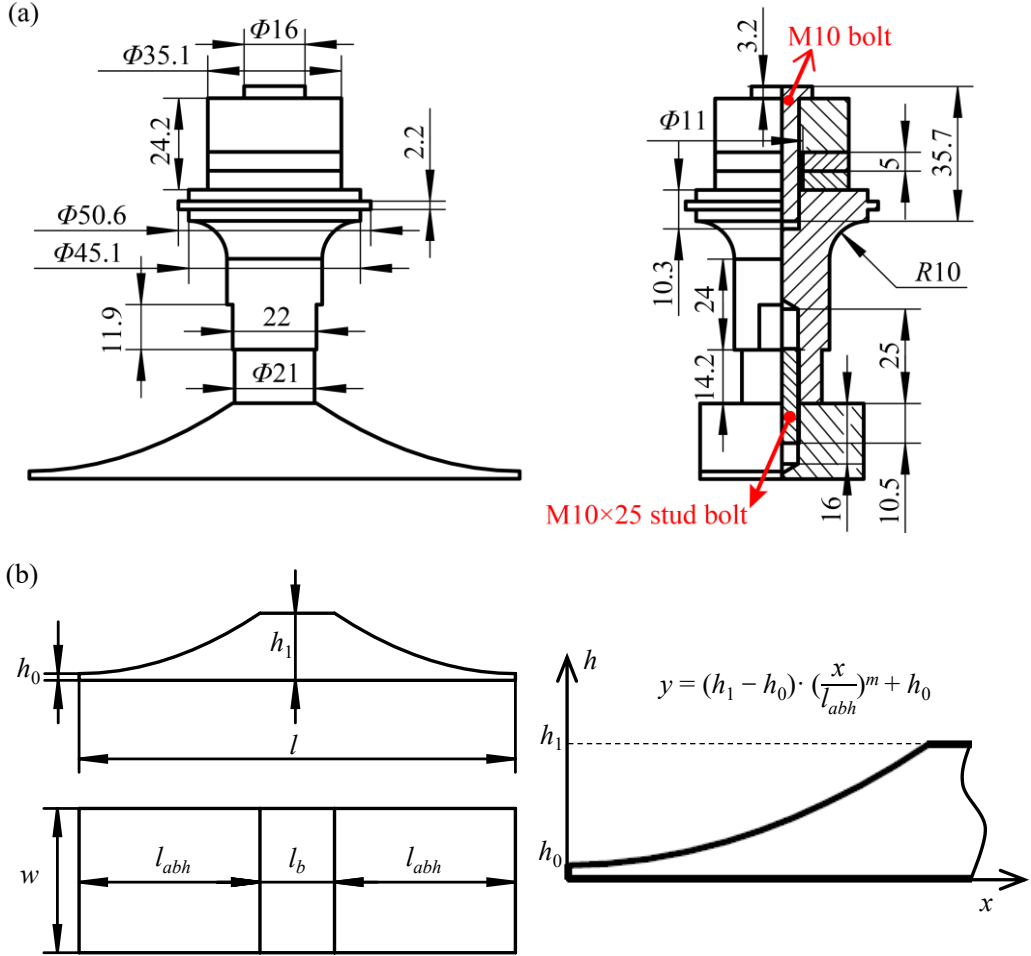

Fig. 2. Schematic diagram of the transducer's dimensions: (a) The Langevin transducer; (b) The ABH wedge radiant plate.

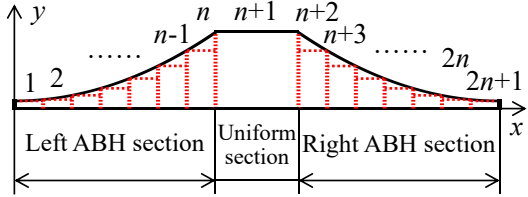

Fig. 3. Division into discrete vibration elements of the radiant plate.

$$Y_b = C_1 \text{ch}(n_1 x) + D_1 \text{sh}(n_1 x) \\ + C_2 \cos(n_2 x) + D_2 \sin(n_2 x), \quad (1)$$

$$Y_s = \Psi_1 [C_1 \text{ch}(n_1 x) + D_1 \text{sh}(n_1 x)] \\ + \Psi_2 [C_2 \cos(n_2 x) + D_2 \sin(n_2 x)]. \quad (2)$$

Here,

$$\Psi_1 = \frac{\omega^2/C_O^2 + n_1^2}{C_S^2 C}, \quad \Psi_2 = \frac{-\omega^2/C_O^2 + n_2^2}{C_S^2 C},$$

$$n_1 = \omega N_1 \sqrt{-1 + N_2 \sqrt{1 + a^2/\omega^2}},$$

$$n_2 = \omega N_1 \sqrt{1 + N_2 \sqrt{1 + a^2/\omega^2}},$$

$$C = \frac{A_0 \rho}{EI}, \quad a = 2\sqrt{C} \bigg/ \left( \frac{1}{C_S^2} - \frac{1}{C_O^2} \right),$$

$$N_1 = \frac{1}{\sqrt{2}} \sqrt{\frac{1}{C_S^2} + \frac{1}{C_O^2}}, \quad N_2 = \frac{C_O^2 - C_S^2}{C_O^2 + C_S^2},$$

$$K' = 5/6, \quad C_O = \sqrt{E/\rho}, \quad C_S = \sqrt{K'G/\rho}.$$

$E$ denotes Young's modulus, $G$ is the shear modulus, $\rho$ represents the density, $I$ is the moment of inertia of the cross-section, $A_0$ is the area of the cross-section, $K'$ is the shear coefficient, $C_O$ is the propagation velocity of longitudinal waves, $C_S$ is the propagation velocity of transverse waves. The remaining parameters are auxiliary variables introduced to simplify the calculation process. Let $Y$, $\Phi$, $M$ and $Q$ denote the total transverse displacement, rotation angle, shear force, and flexural moment, respectively, which are given by

$$Y = Y_b + Y_s, \quad \Phi = \partial V_b / \partial x$$

$$M = EI(\partial^2 V_b / \partial x^2), \quad Q = -K' A_0 G(\partial V_s / \partial x).$$

The vibration parameters of the beam can be expressed as

$$Y = (1+\Psi_1)C_1\mathrm{ch}(n_1 x) + (1+\Psi_1)D_1\mathrm{sh}(n_1 x)$$
$$+(1+\Psi_2)C_2\cos(n_2 x) + (1+\Psi_2)D_2\sin(n_2 x), \quad (3)$$

$$\Phi = C_1 n_1 \mathrm{sh}(n_1 x) + D_1 n_1 \mathrm{ch}(n_1 x)$$
$$- C_2 n_2 \sin(n_2 x) + D_2 n_2 \cos(n_2 x), \quad (4)$$

$$M = EI[C_1 n_1^2 \mathrm{ch}(n_1 x) + D_1 n_1^2 \mathrm{sh}(n_1 x)$$
$$- C_2 n_2^2 \cos(n_2 x) - D_2 n_2^2 \sin(n_2 x)], \quad (5)$$

$$Q = -K'A_0 G\{\Psi_1[C_1 n_1 \mathrm{sh}(n_1 x) + D_1 n_1 \mathrm{ch}(n_1 x)]$$
$$+\Psi_2[-C_2 n_2 \sin(n_2 x) + D_2 n_2 \cos(n_2 x)]\}. \quad (6)$$

As depicted in Fig. 4, the transverse displacement, rotation angle, flexural moment, and shear force of the left and right ends of the $i$-th vibration element ($1 \leq i \leq 2n+1$, $i \in N^*$) are represented by $Y_i$, $\Phi_i$, $M_i$, $Q_i$, $Y_{i+1}$, $\Phi_{i+1}$, $M_{i+1}$, $Q_{i+1}$, respectively. Their relationship is expressed as

$$\begin{bmatrix} Y_i \\ \Phi_i \\ M_i \\ Q_i \end{bmatrix} = \begin{bmatrix} a_{11}^i & a_{12}^i & a_{13}^i & a_{14}^i \\ a_{21}^i & a_{22}^i & a_{23}^i & a_{24}^i \\ a_{31}^i & a_{32}^i & a_{33}^i & a_{34}^i \\ a_{41}^i & a_{42}^i & a_{43}^i & a_{44}^i \end{bmatrix} \begin{bmatrix} Y_{i+1} \\ \Phi_{i+1} \\ M_{i+1} \\ Q_{i+1} \end{bmatrix}, \quad (7)$$

where $a^i$ is the 4×4 matrix for the $i$-th vibration element. Its element, $a_{ij}^i$, depends on the material properties, dimensions, and vibration frequency, as further detailed in the Appendix.

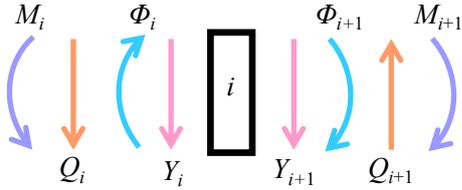

Fig. 4. Boundary mechanical quantities of $i$-th discrete vibration element.

At the interface between any two adjacent vibrating elements, both the displacement and rotation are continuous, and the equilibrium of flexural moment and shear force is satisfied. Therefore, we have

$$\begin{bmatrix} Y_1 \\ \Phi_1 \\ M_1 \\ Q_1 \end{bmatrix} = \begin{bmatrix} A_{11} & A_{12} & A_{13} & A_{14} \\ A_{21} & A_{22} & A_{23} & A_{24} \\ A_{31} & A_{32} & A_{33} & A_{34} \\ A_{41} & A_{42} & A_{43} & A_{44} \end{bmatrix} \begin{bmatrix} Y_{2n+1} \\ \Phi_{2n+1} \\ M_{2n+1} \\ Q_{2n+1} \end{bmatrix}, \quad (8)$$

where $A$ represents the overall matrix of the radiant plate, which can be expressed as

$$A = \prod_{i=1}^{2n+1} a^i. \quad (9)$$

Since the two ends of the radiant plate are free, the flexural moment and shear force are zero, i.e., $M_1 = Q_1 = M_{2n+1} = Q_{2n+1} = 0$. Substituting these conditions into Eq. (8), we obtain

$$\begin{bmatrix} 0 \\ 0 \end{bmatrix} = \begin{bmatrix} A_{31} & A_{32} \\ A_{41} & A_{42} \end{bmatrix} \begin{bmatrix} Y_{2n+1} \\ \Phi_{2n+1} \end{bmatrix}, \quad (10)$$

For Eq. (10) to have a non-zero solution, the determinant of the coefficient matrix must be zero, which can be expressed as

$$\begin{vmatrix} A_{31} & A_{32} \\ A_{41} & A_{42} \end{vmatrix} = A_{31}A_{42} - A_{41}A_{32} = 0. \quad (11)$$

This is the frequency equation, and the frequencies satisfying it represent the flexural vibration frequencies of the radiant plate[24].

## 2.2. Finite-element verification of the theoretical model

Firstly, the geometric parameters of the radiant plate were determined. The length of a single ABH contour, $l_{abh}$, was set to 53.4 mm. The length of the uniform-thickness section, $l_b$, was 22 mm. The total length of the radiant plate, $l$, was 128.8 mm. The thickness of the uniform-thickness part, $h_1$, was 20 mm. The plate width, $w$, was 40 mm, and the truncation thickness, $h_0$, was 2 mm. The steel material used had a density $\rho$ of 7850 kg/m$^3$, a Young's modulus $E$ of 205 GPa, and a Poisson's ratio $\nu$ of 0.28. Subsequently, an FES model of the radiant plate was established. Screw hole and assembly prestress were neglected. The Solid Mechanics module was utilized. As shown in Fig. 5, the maximum mesh element size at both ends of the radiant plate was set to 1.75 mm, while the remaining section of the radiant plate used a size of 3.5 mm. Then, the flexural vibration frequencies of the radiant plate were calculated and compared with the frequencies obtained by solving Eq. (11) using MATLAB R2022a. The results are presented in Table 1, where $f_T$ and $f_F$ denote the flexural vibration frequencies obtained by transfer matrix method (TMM) and FES, respectively. $\Delta_1 = (f_T - f_F)/f_F$ represents the relative error between the results from the TMM and the FES. The results indicate that the theoretical results are in excellent agreement with the FES results, with relative error being less than 4%. This validates the practicability and accuracy of the transfer matrix theoretical model based on Timoshenko beam theory. Furthermore, the theoretical model offers rapid calculation speed and requires fewer computational resources, significantly enhancing the efficiency of the design process. The primary reason for the errors is that the theoretical model only considers one-dimensional flexural vibration, thereby neglecting the coupling effects between different modes.

Table 1 Comparison of calculated flexural vibration frequency of the radiant plate.

| Parameters | 1st order | 2nd order | 3rd order | 4th order | 5th order | 6th order |
|---|---|---|---|---|---|---|
| $f_T$ (Hz) | 5575 | 13909 | 26722 | 43370 | 62585 | 83304 |
| $f_F$ (Hz) | 5431 | 13402 | 26101 | 43287 | 62847 | 83737 |
| $\Delta_1$ (%) | 2.65 | 3.78 | 2.38 | 0.19 | 0.42 | 0.52 |

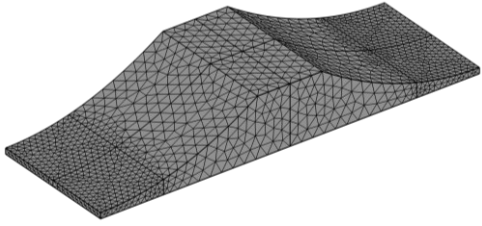

Fig. 5. Meshing of the radiant plate.

## 3. Transducer performance simulations

### 3.1. Simulation of frequency characteristic and vibration mode

As Table 1 shows, the frequency of the third-order symmetrical flexural vibration of the radiant plate largely aligns with the first-order longitudinal vibration frequency of the Langevin transducer. Consequently, this specific mode of the radiant plate was selected for the design of the ABH transducer. The FES model of the ABH transducer was established using COMSOL Multiphysics 6.1. The mesh settings for the radiant plate were consistent with those detailed in Section 2.2 , and the maximum mesh element size of the Langevin transducer was set to 8 mm (Fig. 6). The Solid Mechanics and Electrostatics modules were employed, with an excitation voltage of 1 V. From the impedance frequency response curve and vibration mode (Fig. 7 and Fig. 8), the simulated vibration frequency ($f_{FES}$) of the ABH transducer was found to be 26132 Hz. The relative error between this frequency and the first-order longitudinal vibration frequency of the Langevin transducer (26055 Hz) was only 0.30%. This confirms the feasibility of the design method presented in this paper.

Along the length direction, the central axis of the radiating surface was taken as a cut line. Its normalized lateral displacement distribution curve was then calculated via FES to illustrate the vibration displacement distribution of the entire radiant plate. As Fig. 9 demonstrates, the amplitude gradually increases from the center towards both ends, reaching its maximum at the ends. This effectively verifies the energy concentration effect of the ABH structure.

### 3.2. Simulation of analysis of near-field sound field

To investigate the near-field sound field of the transducer in air, the Pressure Acoustic module in COMSOL was used (Fig. 10). A hemisphere with a radius of 110 mm was added as the air domain, and a perfectly matched layer (PML), with a thickness of 15 mm, was applied to its outer surface to absorb incident sound waves. To ensure the simulation accuracy of acoustic waves in air, the maximum mesh element size of the air domain was set to 1/6 of the corresponding wavelength. Subsequently, the PML was divided into five mesh layers, and a boundary layer with a thickness of 0.01 times the corresponding wavelength was created at the interface with the air domain, ensuring full absorption of incident sound waves.

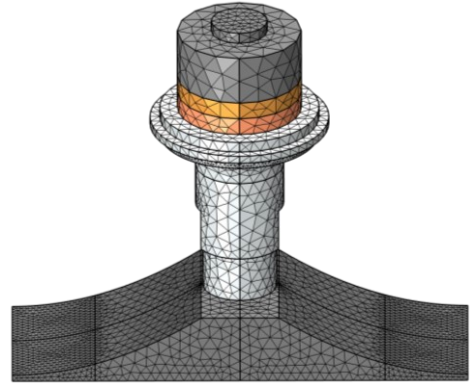

Fig. 6. Meshing of the entire transducer.

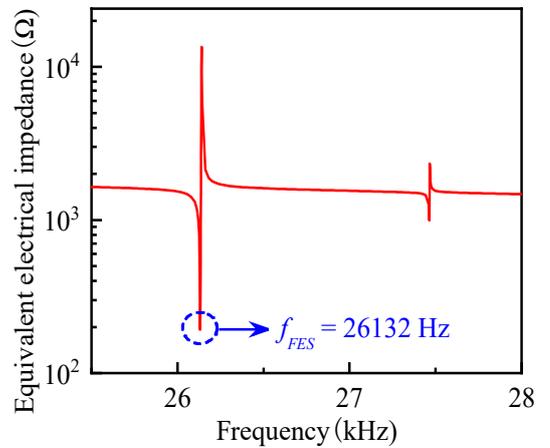

Fig. 7. Electrical impedance frequency response curve of the transducer obtained from FES.

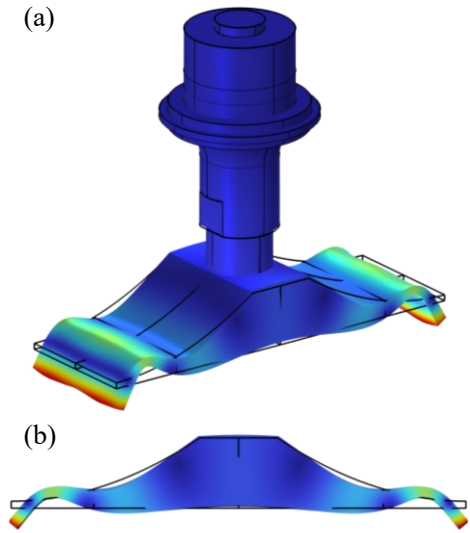

Fig. 8. (a) Vibration mode of the entire transducer; (b) the flexural vibration mode of the radiant plate.

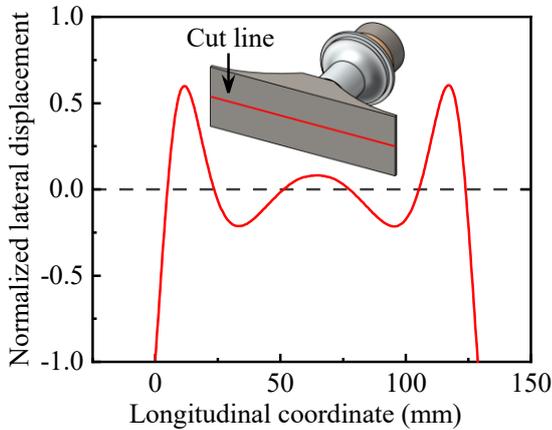

Fig. 9. Curve of normalized transverse displacement distribution of the radiant plate.

The excitation voltage was set to 1 V, and the excitation frequency to 26132 Hz. As Fig. 11 shows, the sound pressure exhibited a gradient distribution to some extent, with the maximum sound pressure occurring at both ends of the radiant plate where the amplitude was largest. Furthermore, diverse energy potential wells were observed in front of the radiant plate. This non-uniform near-field sound pressure distribution holds significant promise for achieving gradient levitation of particles.

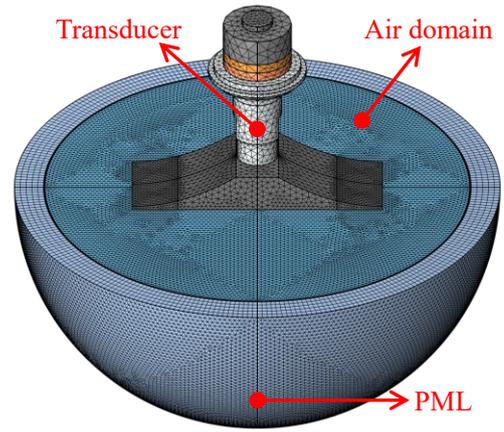

Fig. 10. Meshing of the near sound field domain in air.

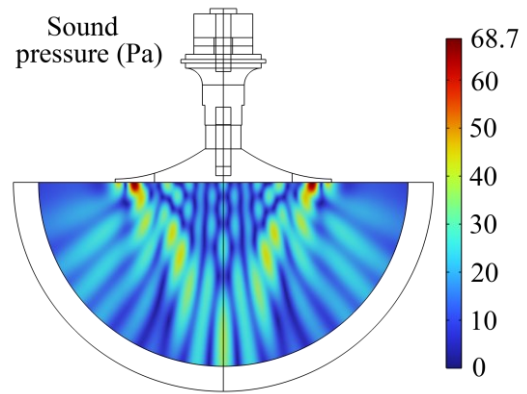

Fig. 11. Near-field sound pressure distribution in air obtained from FES.

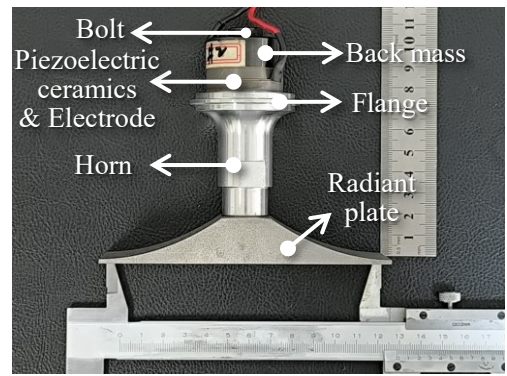

Fig. 12. Prototype of the ultrasonic transducer with ABH wedge structure based on vibration mode-conversion.

## 4. Experimental verification

To verify the feasibility of the ABH transducer design, a transducer prototype, as shown in Fig. 12, was fabricated. The bolt, back mass, and radiant plate were made of steel; the horn was made of aluminum; the piezoelectric ceramic was PZT-4; and the electrodes were made of copper. An impedance analyzer and a laser vibrometer were employed to measure the electrical parameters and vibration modes of the transducer prototype. Additionally, a microphone was used to measure the near-field radiated sound pressure of the prototype in air. Finally, the ultrasonic levitation experiment was conducted.

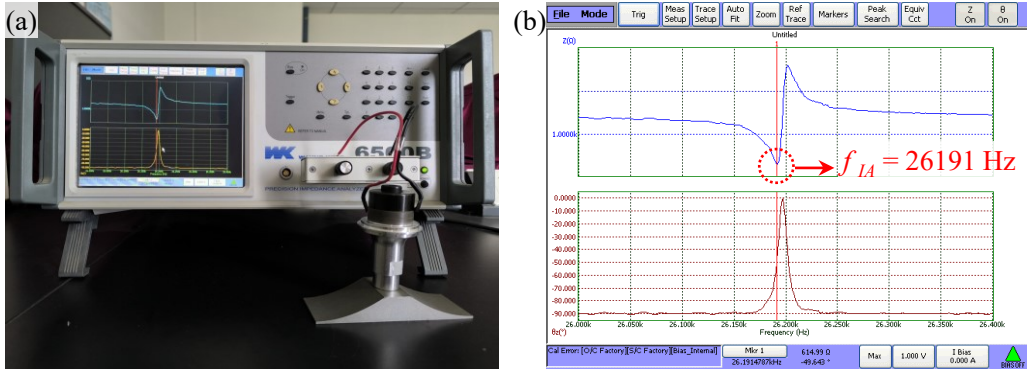

Fig. 13. Impedance analysis experiment: (a) The precision impedance analyzer and the transducer prototype; (b) Frequency response curve of equivalent electrical impedance and phase of the transducer.

### 4.1. Experiment of frequency characteristic measurement

To validate the FES results, the electrical impedance of the transducer was measured using a precision impedance analyzer ((WK6500B, Wayne Kerr Electronics, UK)), as shown in Fig. 13(a). The excitation voltage set to 1 V. As illustrated by the electrical impedance frequency response curve in Fig. 13(b), the resonant frequency of the transducer prototype, $f_{IA}$, was 26191 Hz. The relative error between this value and the resonant frequency obtained from FES, $f_{FES}$, was calculated as $\Delta_2 = (f_{IA} - f_{FEM})/f_{IA} = 0.23\%$, indicating good agreement and verifying the reliability of the FES frequency results. The main sources of error are attributed to the following three factors:

1) Deviations between the actual material parameters of the fabricated prototype and the standard material parameters used in the FES;
2) The FES ignoring the mechanical and dielectric losses of the transducer, as well as the prestress induced by the bolts;
3) Limitations in the machining accuracy of the prototype and assembly errors in the transducer.

### 4.2. Experiment of vibration mode measurement

The vibration mode of the transducer was measured using a full-field scanning vibrometry system (PSV-400, Polytec, Germany), as shown in Fig. 14(a). The excitation voltage was set to 1 V, and the frequency sweep range was 20-30 kHz. As Fig. 14(b) shows, the experimentally measured vibration mode demonstrated good agreement with those obtained from FES, thereby confirming that the target mode could be effectively excited.

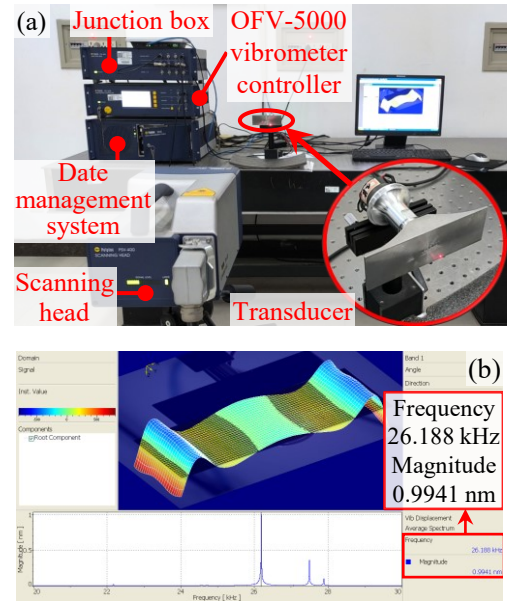

Fig. 14. Laser vibrometry experiment: (a) The full-field scanning vibrometry system and the transducer prototype; (b) Flexural vibration mode of the radiant plate.

### 4.3. Experiment of near-field sound pressure measurement

The sound field measurement system was set up as shown in Fig. 15. First, the transducer was electrically excited: a sinusoidal electrical signal of 26161 Hz and 500 mVpp was generated by a signal generator (DG1022U, RIGOL Technologies, China), which was amplified by a power amplifier (AG1012, T&C Power Conversion, USA) to yield an effective output power of 4 W. Subsequently, the transducer was electrically matched using an inductance matching box (AI 7-1 Adjustable Inductor, T&C Power Conversion, USA) and an impedance matching box (MODEL SUT 2K LF-7, T&C Power Conversion, USA) for optimal excitation. A microphone (Type 3052-A-030, Brüel & Kjær, Denmark) was placed directly in front of the radiating surface and connected to a data acquisition hardware (3052-A-030, Brüel & Kjær, Denmark), with settings to average the data acquired over 20 s. The voltage across the transducer was measured using a digital multimeter (34401A, Agilent Technologies, USA).

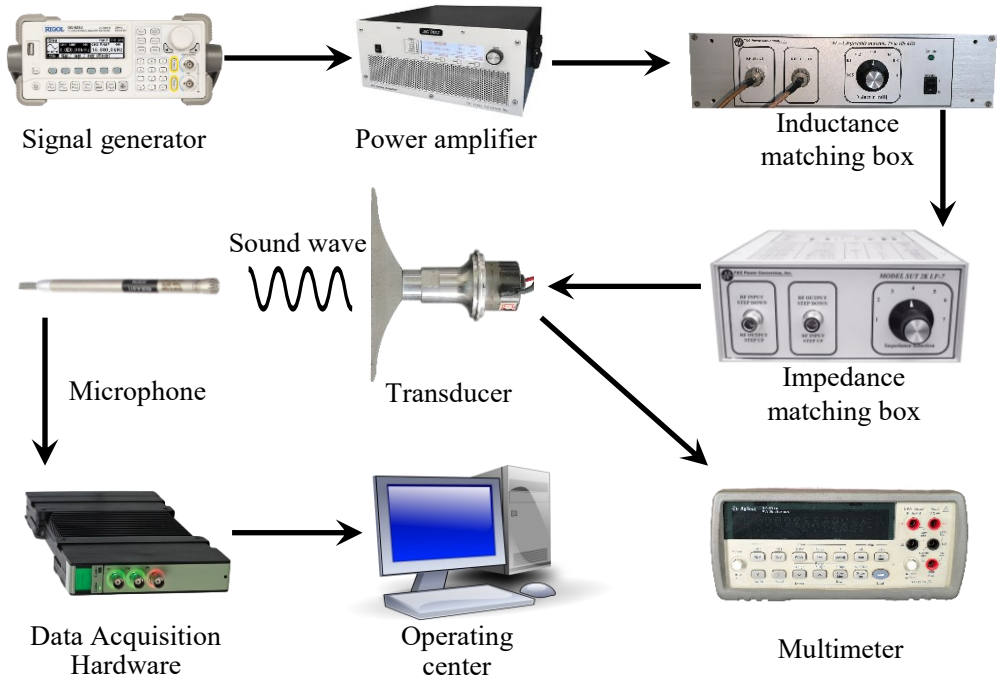

Fig. 15. Experimental setup of acoustic field measurement in air.

As Fig. 16(a) illustrates, the near-field radiated sound pressure in air was measured on a plane 10 mm away from the radiating surface, specifically along the length, width, and axial directions of the transducer. Owing to the symmetry of the near sound field distribution, 14 measurement points were uniformly taken on one side along the length direction, and 5 measurement points were uniformly taken on one side along the width direction. Furthermore, 22 measuring points were uniformly acquired along the axial direction. As indicated in Figs. 16(b), (c) and (d), the near-field sound pressure distribution obtained from the FES and experimental measurement showed good consistency. Deviations in some data points were primarily attributed to sound wave reflections and refractions during the experiment. The near-field sound pressure distribution plots reveal that, axially, the near-field sound pressure amplitude fluctuates significantly; as the distance increases, the spacing between maximum and minimum sound pressures gradually expands, and the maximum values progressively increase. In the width direction, the sound pressure exhibited a characteristic distribution with higher pressure in the middle and lower pressure on both sides. Along the length direction, the sound pressure gradually intensified from the center towards both sides, with the maximum sound pressure appearing at the two ends. This phenomenon is attributed to the energy concentration effect of the ABH structure, which progressively amplifies the vibration amplitude at both sides of the radiant plate. This leads to stronger radiated sound pressure closer to the ends, consequently resulting in the radiant plate's near-field sound pressure distribution exhibiting a gradient increase. This characteristic can be leveraged for achieving gradient levitation of particles.

### 4.4. Experiment of ultrasonic levitation

Fig. 17(a) illustrates the constructed ultrasonic levitation experimental platform. The radiant surface of the transducer was oriented downwards, with a cubic aluminum block placed beneath it to serve as a reflection baffle, thereby forming a standing wave sound field between the transducer and the aluminum block. The signal generator was set to generate a sinusoidal electrical signal of 26161 Hz and 500 mVpp, and the power amplifier delivered an effective output power of 4 W. Several expanded polystyrene (EPS) spheres, each with a diameter of 4 mm, were then levitated in the sound field. As shown in Fig. 17(b), when the radiant surface and the reflection baffle were separated by approximately 20.3 mm, 5, 4, 3, 4, and 5 EPS spheres were observed levitated sequentially from left to right. This observation indicates that, due to the cumulative amplification effect of the ABH structure on the flexural amplitude, the standing wave sound field between the radiant surface and the reflection baffle exhibits a gradient distribution along the length direction. Consequently, the acoustic radiation force gradually increases from the middle towards both sides, leading to a corresponding increase in the number of levitated spheres. This suggests the potential for achieving particle sorting.

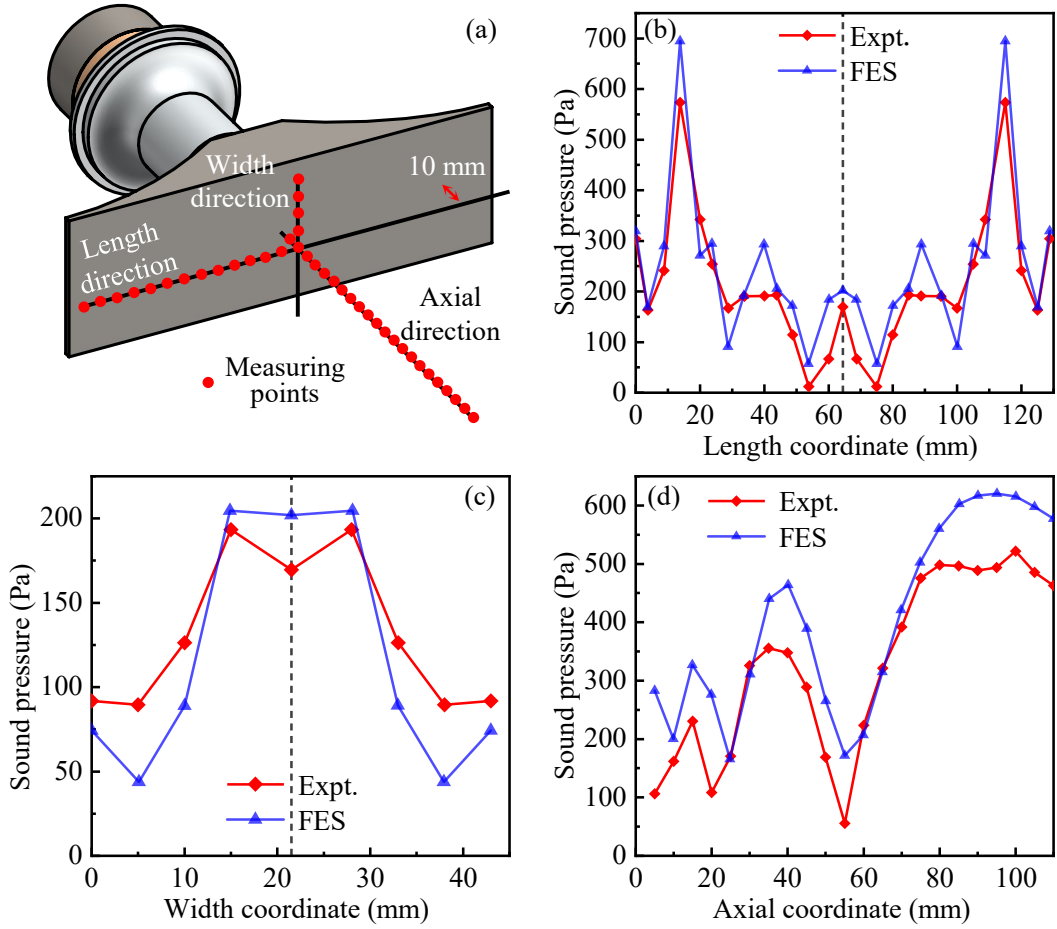

Fig. 16. Comparison of the experimental and FES's results of near-field sound pressure distribution in air: (a) Schematic diagram of measuring points; (b) Sound pressure distribution in the length direction 10 mm away from the radiant surface; (c) Sound pressure distribution in the width direction 10 mm away from the radiant surface; (d) Axial sound pressure distribution.

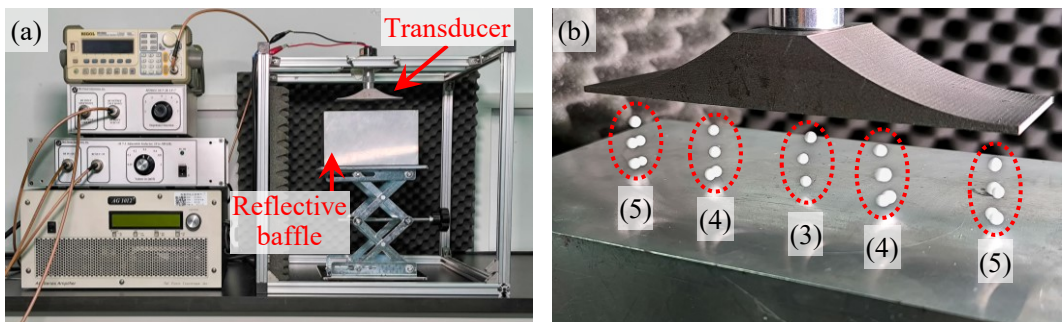

Fig. 17. (a) Platform of ultrasonic levitation experiment; (b) gradient levitation of EPS spheres.

## 5. Conclusion

In this paper, we designed a vibration mode-conversion ultrasonic transducer incorporating an ABH wedge structure, leveraging the energy concentration effect of the ABH structure. A theoretical model, based on the Timoshenko beam theory and the transfer matrix method, was established. This model enables the rapid and accurate calculation of the radiant plate's flexural vibration frequencies, significantly enhancing the design efficiency of the radiant plate. According to the results from both FES and experimental measurement, the amplitude of the radiant plate progressively increases from the center towards both ends. This phenomenon leads to a gradient distribution characteristic in the near-field radiated sound pressure, with diverse potential wells observed within the sound field. The experimentally measured vibration frequency, mode, and near-field sound pressure distribution showed good agreement with the FES results, thereby validating the feasibility of the design method presented in this paper. The particle gradient levitation experiment confirmed the potential of the ABH transducer to construct a multi-functional sound field, suggesting promising future applications in areas such as particle sorting. Looking ahead, we will continue to optimize the engineering design of ABH structures and explore their application potential in multifunctional particle manipulation.

## Acknowledgments

This work is supported by the National Natural Science Foundation of China [Grant Nos. 12174240, 11874253 and 11674206].

## Appendix

Given that each element $a_{mn}^i$ in Eq. (7) of the main text is a 4×4 matrix, its elements are as follows:

$$a_{11} = \frac{(1+\Psi_1)n_2^2 \text{ch}(n_1 l) + (1+\Psi_2)n_1^2 \cos(n_2 l)}{n_1^2(1+\Psi_2) + n_2^2(1+\Psi_1)}, \quad (A1)$$

$$a_{12} = \frac{n_1\Psi_1(1+\Psi_2)\sin(n_2 l) - n_2\Psi_2(1+\Psi_1)\text{sh}(n_1 l)}{n_1 n_2(\Psi_2 - \Psi_1)}, \quad (A2)$$

$$a_{13} = \frac{(1+\Psi_1)(1+\Psi_2)[\text{ch}(n_1 l) - \cos(n_2 l)]}{EI[n_1^2(1+\Psi_2) + n_2^2(1+\Psi_1)]}, \quad (A3)$$

$$a_{14} = \frac{(1+\Psi_2)n_1 \sin(n_2 l) + (1+\Psi_1)n_2 \text{sh}(n_1 l)}{K'A_0 G n_1 n_2(\Psi_2 - \Psi_1)}, \quad (A4)$$

$$a_{21} = \frac{n_2 n_1^2 \sin(n_2 l) - n_1 n_2^2 \text{sh} n_1 l}{n_1^2(1+\Psi_2) + n_2^2(1+\Psi_1)}, \quad (A5)$$

$$a_{22} = \frac{\Psi_2 \text{ch}(n_1 l) - \Psi_1 \cos(n_2 l)}{\Psi_2 - \Psi_1}, \quad (A6)$$

$$a_{23} = \frac{-n_1(1+\Psi_2)\text{sh}(n_1 l) - n_2(1+\Psi_1)\sin(n_2 l)}{EI[n_1^2(1+\Psi_2) + n_2^2(1+\Psi_1)]}, \quad (A7)$$

$$a_{24} = \frac{\text{ch}(n_1 l) - \cos(n_2 l)}{K'A_0 G(\Psi_2 - \Psi_1)}, \quad (A8)$$

$$a_{31} = \frac{EI n_1^2 n_2^2 [\text{ch}(n_1 l) - \cos(n_2 l)]}{n_1^2(1+\Psi_2) + n_2^2(1+\Psi_1)}, \quad (A9)$$

$$a_{32} = \frac{-EI[\Psi_2 n_1 \text{sh}(n_1 l) + \Psi_1 n_2 \sin(n_2 l)]}{\Psi_2 - \Psi_1}, \quad (A10)$$

$$a_{33} = \frac{n_1^2(1+\Psi_2)\text{ch}(n_1 l) + n_2^2(1+\Psi_1)\cos(n_2 l)}{n_1^2(1+\Psi_2) + n_2^2(1+\Psi_1)}, \quad (A11)$$

$$a_{34} = \frac{-EI[n_1 \text{sh}(n_1 l) + n_2 \sin(n_2 l)]}{K'A_0 G(\Psi_2 - \Psi_1)}, \quad (A12)$$

$$a_{41} = \frac{K'A_0 G[n_2^2 n_1 \Psi_1 \text{sh}(n_1 l) - n_1^2 n_2 \Psi_2 \sin(n_2 l)]}{n_1^2(1+\Psi_2) + n_2^2(1+\Psi_1)}, \quad (A13)$$

$$a_{42} = \frac{K'A_0 G[\cos(n_2 l) - \text{ch}(n_1 l)]\Psi_1 \Psi_2}{\Psi_2 - \Psi_1}, \quad (A14)$$

$$a_{43} = \frac{K'A_0 G[n_1 \Psi_1(1+\Psi_2)\text{sh}(n_1 l) + n_2 \Psi_2(1+\Psi_1)\sin(n_2 l)]}{EI[n_1^2(1+\Psi_2) + n_2^2(1+\Psi_1)]}, \quad (A15)$$

$$a_{44} = \frac{\Psi_2 \cos n_2 l + \Psi_1 \text{ch} n_1 l}{\Psi_2 - \Psi_1}. \quad (A16)$$

Here,

$$\Psi_1 = \frac{\omega^2/C_O^2 + n_1^2}{C_S^2 C}, \quad \Psi_2 = \frac{-\omega^2/C_O^2 + n_2^2}{C_S^2 C},$$

$$n_1 = \omega N_1 \sqrt{-1 + N_2 \sqrt{1 + a^2/\omega^2}},$$

$$n_2 = \omega N_1 \sqrt{1 + N_2 \sqrt{1 + a^2/\omega^2}},$$

$$C = \frac{A_0 \rho}{EI}, \quad a = 2\sqrt{C} \bigg/ \left(\frac{1}{C_S^2} - \frac{1}{C_O^2}\right),$$

$$N_1 = \frac{1}{\sqrt{2}}\sqrt{\frac{1}{C_S^2}+\frac{1}{C_O^2}},\quad N_2 = \frac{C_O^2-C_S^2}{C_O^2+C_S^2},\quad K'=5/6,\quad C_O=\sqrt{E/\rho},\quad C_S=\sqrt{K'G/\rho}.$$